\documentclass[11pt]{article}

\usepackage{graphicx}
\usepackage{amsmath}
\usepackage{dcolumn}
\usepackage{bm}
\usepackage[utf8]{inputenc}
\usepackage{authblk}
\usepackage{cite}

\newcommand{\be}{\begin{equation}}
\newcommand{\ee}{\end{equation}}
\newcommand{\ba}{\begin{eqnarray}}
\newcommand{\ea}{\end{eqnarray}}
\newcommand{\no}{\nonumber \\}
\newcommand{\gsim}{\mathrel{\hbox{\rlap{\lower.55ex \hbox {$\sim$}}
                   \kern-.3em \raise.4ex \hbox{$>$}}}}
\newcommand{\lsim}{\mathrel{\hbox{\rlap{\lower.55ex \hbox {$\sim$}}
                   \kern-.3em \raise.4ex \hbox{$<$}}}}

\def\roughly#1{\mathrel{\raise.3ex\hbox{$#1$\kern-.75em%
\lower1ex\hbox{$\sim$}}}}
\def\lsim{\roughly<}
\def\gsim{\roughly>}

\def\({\left(}
\def\){\right)}
\def\[{\left[}
\def\]{\right]}
\def\<{\langle}
\def\>{\rangle}
\def\pd{\partial}

\def\l{{\lambda}}
\def\L{{\Lambda}}
\def\d{{\delta}}
\def\D{{\Delta}}
\def\o{{\omega}}

\def\e{{\epsilon}}

\def\G{{\Gamma}}

\def\n{{\nu}}

\def\t{{\tau}}
\def\tht{{\theta}}

\def\x{{\xi}}

\newcommand{\Z}{\int_0^z\frac{dz'}{f(z')}}
\newcommand{\hph}{{\hat\phi}}
\newcommand{\ads}{\text{AdS}}
\newcommand{\btz}{\text{BTZ}}
\newcommand{\thT}{\text{th}}
\newcommand{\oar}{\overleftrightarrow}

\title{\bf Holographic thermalization with initial long range correlation}

\author[1,2]{Shu Lin\thanks{linshu8@mail.sysu.edu.cn}}
\affil[1]{Institute of Astronomy and Space Science, Sun Yat-Sen University, No 135 Xingang Xi Rd, Guangzhou, 510275, China}
\affil[2]{RIKEN-BNL Research Center, Brookhaven National Laboratory, Upton, NY 11973, USA}

\date{\today}

\begin{document}

\maketitle

\begin{abstract}
We studied the evolution of Wightman correlator in a thermalizing state modeled by AdS${}_3$-Vaidya background. We gave a prescription for calculating Wightman correlator in coordinate space without using any approximation. For equal-time correlator $\<O(v,x)O(v,0)\>$, we obtained an enhancement factor $v^2$ due to long range correlation present in the initial state. This was missed by previous studies based on geodesic approximation. We found that the long range correlation in initial state does not lead to significant modification to thermalization time as compared to known results with generic initial state. We also studied spatially integrated Wightman correlator and showed evidence on the distinction between long distance and small momentum physics for an out-of-equilibrium state. We also calculated radiation spectrum of particle weakly coupled to $O$ and found lower frequency mode approaches thermal spectrum faster than high frequency mode.
\end{abstract}

\newpage

\section{Introduction and Summary}

The phenomenon of thermalization, where a quantum field state evolves unitarily from a pure state to an apparent thermal state, exists in many different areas of physics, including heavy ion collisions, cold atom system etc. Since systems having thermalization phenomenon are usually strongly coupled, theoretical studies of thermalization remain a difficult task. The application of gauge/gravity duality allows us to study dynamics of strongly coupled field theory by solving weakly coupled gravity in one dimension higher, providing a useful alternative to traditional methods.
Recently, there have been significant progress in understanding thermalization using holographic method.

Useful probes of thermalization process are correlation functions. These include local probes like one-point function and nonlocal probes like two-point function. They contain different information on the thermalization process. For example, in the context of heavy ion collisions, one-point function of stress energy tensor determines spectrum of hadron, which is emitted only at freezeout, while two-point function of electromagnetic current determines spectrum of photon/dilepton, which is being emitted throughout the history of quark gluon plasma evolution. 
The calculation of one-point functions, usually involving solving Einstein equations, has been pursued by many groups, see for example \cite{Bhattacharyya:2009uu,Grumiller:2008va,Chesler:2015wra,Chesler:2013lia,Chesler:2010bi,Chesler:2009cy,Chesler:2008hg,Bantilan:2012vu,Beuf:2009cx,Heller:2011ju,vanderSchee:2013pia,Casalderrey-Solana:2013aba,Garfinkle:2011tc,Garfinkle:2011hm,Wu:2011yd,Romatschke:2013re,Caceres:2014pda}. We will focus on two-point functions in this work. The calculation of two-point functions, as pointed out in \cite{CaronHuot:2011dr}, depends on the order of operators. While the retarded correlator is independent of field theory state, the Wightman correlator ``does'' depend on state. Consequently, the former can be obtained by studying response of bulk field to external boundary source. On the contrary, the calculation of the latter must be formulated as an initial value problem, with initial value encoding state information.

Previous studies on two-point function used different approximation schemes, such as quasi-static approximation \cite{Danielsson:1999fa,Danielsson:1999zt,Giddings:2001ii,Lin:2008rw,Baron:2012fv,Steineder:2012si,Steineder:2013ana,Baier:2012ax,Baier:2012tc}, geodesic approximation \cite{Aparicio:2011zy,Balasubramanian:2011ur,Balasubramanian:2010ce,Balasubramanian:2013oga,Balasubramanian:2013rva,Balasubramanian:2012tu,Galante:2012pv,Caceres:2012em}, geometric optics approximation \cite{Hubeny:2006yu,Erdmenger:2011jb,Erdmenger:2012xu,Erdmenger:2011aa,Chesler:2012zk,Chesler:2011ds,CaronHuot:2011dr} etc. Recently, a rigorous calculation has been done by Ker\"anen and Kleinert (KK) for Wightman correlator in (spatial) momentum space \cite{Keranen:2014lna}, see also \cite{Jarvela:2015zra,Ebrahim:2010ra,David:2015xqa} for related attempts. In this work, we will present results for the same correlator in coordinate space. For technical reasons, we focus on AdS$_{3}$-Vaidya or thermalization of $1+1$D conformal field theory (CFT). Our results in coordinate space allow for direct comparison to general CFT results by Calabrese and Cardy (CC) \cite{Calabrese:2006rx}. While CC assumes finite correlation length in the initial state, our initial (vacuum) state contains long range correlation. Consequently our results show reminiscent of long range correlation: The equal-time correlator $\<O(v,x)O(v,0)\>$ for a scalar operator $O$ with dimension $2$ has power law factor $v^2/x^4$ in the regime $x\gg v\gg 1$. Previous studies based on geodesic approximation capture the $x$-dependence, but missed the $v$-dependence.
The general results of CC determine a thermalization time $x+O(1/T)$ for points with spatial separation $x$, and a thermalization time $O(1/T)$ for points with temporal separation $\d v$. We confirmed that the conclusion is still true when initial state has long range correlation.

We have also studied spatially integrated correlator, that is zero (spatial) momentum mode of Wightman correlator. While low momentum mode is usually regarded as equivalent to long distance physics. We found a counter example: when the state is far from equilibrium, the zero momentum correlator receives dominant contribution from integrating over correlator with small separation, i.e. short distance physics. Furthermore, we calculate the spectrum of particle weakly coupled to operator $O$ at different stage of thermalization. Our results indicate that high frequency modes tend to appear thermal slower than low frequency modes.

\section{Probing a thermalization process with Wightman correlator}

We consider a thermalizing state modeled by AdS${}_3$-Vaidya background below
\begin{align}\label{metric}
ds^2=\frac{-fdv^2-2dvdz+dx^2}{z^2},
\end{align}
with $f=1-mz^2\tht(v)$. Here $v$ is a lightcone coordinate, which is related to usual coordinates $t$ and $z$ by
\begin{align}\label{v_def}
dv=dt-\Z.
\end{align}
The metric corresponds to a lightlike shell (shock wave) collapsing from the boundary at $v=0$. The hypersurface $v=0, x=\text{arbitrary}$ separates empty AdS and BTZ metrics on different sides of the shell.
Empty AdS and BTZ metrics are dual to vacuum and thermal CFT states respectively. The Vaidya background is dual to a thermalization process, triggered by an instantaneous injection of energy density to vacuum state at $t=0$. The end point of the thermalization process is a thermal state with temperature $T=\sqrt{m}/2\pi$. We will set $m=1$ from now.
%
We choose to probe the thermalizing state by Wightman correlator $\<O(t,x)O(t',x')\>$. $O$ is a dimension $2$ scalar operator dual to bulk dilaton.
Unlike retarded correlator, Wightman correlator cannot be calculated from the response of bulk field to source on the boundary. This is because the retarded correlator is state independent, while the Wightman correlator depends on the state. In \cite{CaronHuot:2011dr}, the calculation of generic correlator was formulated as an initial value problem. The equivalence of the formulation with standard holographic dictionary was shown by KK \cite{Keranen:2014lna}, based on more general prescription for real-time holography \cite{Skenderis:2008dh,Skenderis:2008dg}. We will use the formulation for our specific setup. The bulk Wightman correlator can be written as
\begin{align}\label{W_rep}
G^>(4|3)=\int dz_1dx_1dz_2dx_2G^>_0(2|1)\oar{D}^{v1}\oar{D}^{v2}G_\thT^R(3|1)G_\thT^R(4|2),
\end{align}
where
$iG^>_0(2|1)=\<\hph(v_2,x_2,z_2)\hph(v_1,x_1,z_1)\>$ is the bulk Wightman correlator evaluated on the hypersurface of the shell $v_2=v_1=0$. This is our initial value. Assuming the continuity of the bulk correlator across the shell, we can use the value of $G^>_0(2|1)$ in empty AdS.
$iG_\thT^R(3|1)=\<[\hph(v_3,x_3,z_3), \hph(v_1,x_1,z_1)]\>\tht(t_3-t_1)$ and similarly for $iG_\thT^R(4|2)$
are retarded bulk-bulk propagators in BTZ, which propagate points $1$ and $2$ from empty AdS to points $3$ and $4$ on BTZ side. The resulting bulk correlator $G^>(4|3)=i\<\hph(v_4,x_4,z_4)\hph(v_3,x_3,z_3)\>$ gives us the boundary Wightman correlator:
\begin{align}
\<G^>(4|3)\>\to z_4^2z_3^2\<O(v_4,x_4)O(v_3,x_3)\>, \quad \text{as}\quad z_4,z_3\to 0.
\end{align}
The symbol $\oar{D}^v=\sqrt{-g}g^{vz}\oar{\pd}_z$ is a two-way differential operator. Note that it only involves derivative with respect to $z$. It means that we need only one initial value $G^>_0(2|1)$. This is in contrast to conventional initial value problems where we need both position and velocity. The reason is that we are using lightcone coordinate $v$ and our initial value hypersurface is also lightlike. The simplification comes with a price: The initial value $G^>_0(2|1)$ is singular as the points $2$ and $1$ approaches the lightcone. Nevertheless, we can eliminate the singularity by subtracting the same quantity evaluated in BTZ space: $iG^>_\thT(2|1)=\<\hph(v_4,x_4,z_4)\hph(v_3,x_3,z_3)\>|_\btz$. Applying \eqref{W_rep} to BTZ background with fictitious hypersurfaces $v_2=v_1=0$, we obtain
\begin{align}\label{Wth_rep}
G_\thT^>(4|3)=\int dz_1dx_1dz_2dx_2G^>_\thT(2|1)\oar{D}^{v1}\oar{D}^{v2}G_\thT^R(3|1)G_\thT^R(4|2).
\end{align}
Subtracting \eqref{Wth_rep} from \eqref{W_rep} and noting that $D^v$ is the same for AdS and BTZ spaces, we obtain
\begin{align}\label{DW_rep}
\D G^>(4|3)=\int dz_1dx_1dz_2dx_2\D G^>(2|1)\oar{D}^{v1}\oar{D}^{v2}G_\thT^R(3|1)G_\thT^R(4|2),
\end{align}
with $\D G^>(4|3)=G^>(4|3)-G_\thT^>(4|3)$ being the difference between bulk correlators in Vaidya background and BTZ background and $\D G^>(2|1)=G_0^>(2|1)-G_\thT^>(2|1)$ being the initial value for $\D G^>(4|3)$. Below we will show $\D G^>(2|1)$ is free of singularity.
It is useful to note that AdS$_3$ ad BTZ metrics are related by coordinate transformation.
\begin{align}\label{ads_btz}
&ds^2_{\ads}=\frac{1}{z^2}\(-(1-z^2)dt^2+\frac{dz^2}{1-z^2}+dx^2\), \no
&ds^2_{\btz}=\frac{1}{\bar{z}^2}\(-d\bar{t}^2+d\bar{z}^2+d\bar{x}^2\).
\end{align}
The explicit coordinate transformation is given by
\begin{align}\label{coord_trans}
&\bar{x}=\sqrt{1-z^2}e^x\cosh t, \no
&\bar{t}=\sqrt{1-z^2}e^x\sinh t, \no
&\bar{z}=ze^x.
\end{align}
The bulk-bulk correlator in Euclidean-AdS is known \cite{D'Hoker:2002aw}
\begin{align}
G_E(2|1)=\frac{2^{-\D}C_\D}{2\D-d}\x^\D F(\frac{\D}{2},\frac{\D+1}{2};\D-\frac{d}{2}+1,\x^2).
\end{align}
For our case of interest $\D=d=2$, $G_E$ is reduced to
\begin{align}
G_E(2|1)=\frac{1}{4\pi}\(\frac{1}{\sqrt{1-\x^2}}-1\).
\end{align}
Using analytic continuation and the coordinate transformation, we obtain the following
\begin{align}\label{ana_cont}
&iG_0^>(2|1)=G_E(\x=\frac{2z_2z_1}{z_2^2+z_1^2-(v_2-v_1+z_2-z_1-i\e)^2+(x_{21})^2}), \no
&iG_\thT^>(2|1)=G_E(\x=\frac{z_2z_1}{\cosh(x_{21})-\sqrt{1-z_2^2}\sqrt{1-z_1^2}\cosh(v_2-v_1+y_2-y_1-i\e)}),
\end{align}
with $y_i=-\frac{1}{2}\ln\frac{1-z_i}{1+z_i},\; i=1,2$. We have also used the short hand notation $x_{ij}=x_i-x_j$.
From \eqref{ana_cont}, we find both $G_0^>$ and $G_\thT^>$ have singularities as $v_2,v_1\to 0$ and $x_{21}\to 0$. In this case $\x\to 1$ and $i\e$-prescription become relevant. The singularities arises when the two points pinch $ds^2\to 0$ (according to \eqref{metric}) thus is owing to short distance physics. Indeed the singularity disappears in the difference $\D G^>(2|1)$, which allows us to drop the $i\e$-prescription:
\begin{align}
i\D G^>(2|1)=G_E(\x=\frac{2z_2z_1}{2z_2z_1+(x_{21})^2})-G_E(\x=\frac{z_2z_1}{\cosh(x_{21})-1+z_2z_1}).
\end{align}
By taking a different limit $z_1\to 0$, $z_2\to 0$ of \eqref{ana_cont}, we can also obtain the Wightman correlators in vacuum and thermal state of $1+1$D CFT:
\begin{align}\label{cft_wight}
&iG^>_0(2|1)=\frac{2}{\pi}\frac{1}{\(-(v_2-v_1-i\e)+x_{21}\)^2}, \no
&iG^>_\thT(2|1)=\frac{1}{2\pi}\frac{1}{\(-\cosh(v_2-v_1-i\e)+\cosh x_{21}\)^2}.
\end{align}
Now we turn to the propagators $G^R_\thT(3|1)$ and $G^R_\thT(4|2)$. To be specific, we discuss $G^R_\thT(3|1)$ as an example. It is given by
\begin{align}\label{GR_th}
iG^R_\thT(3|1)&=i\(G^>_\thT(3|1)-G^<_\thT(3|1)\)\tht(t_3-t_1) \no
&=\bigg[G_E(\x=\frac{z_3z_1}{\cosh(x_{31})-\sqrt{1-z_3^2}\sqrt{1-z_1^2}\cosh(v_3-v_1+y_3-y_1-i\e)^2}) \no
&-G_E(\x=\frac{z_3z_1}{\cosh(x_{31})-\sqrt{1-z_3^2}\sqrt{1-z_1^2}\cosh(v_3-v_1+y_3-y_1+i\e)^2})\bigg]\tht(t_3-t_1)
\end{align}
In the limit $z_3\to 0$ that we are interested in the calculation of boundary correlator, the propagator simplifies:
\begin{align}\label{GR_zto0}
G^R_\thT(3|1)&=-\frac{i}{8\pi}z_3^2\bigg[\frac{z_1^2}{(\cosh(x_{31})-\sqrt{1-z_1^2}\cosh(v_3-y_1-i\e))^2} \no
&-\frac{z_1^2}{(\cosh(x_{31})-\sqrt{1-z_1^2}\cosh(v_3-y_1+i\e))^2}\bigg]\tht(t_3-t_1),
\end{align}
where we have used $v_1=0$.
Note that the propagator is only nonvanishing on the lightcone:
\begin{align}\label{lightcone}
\cosh(x_{31})=\sqrt{1-z_1^2}\cosh(v_3-y_1).
\end{align}
Since $\cosh(x_{31})\ge1$, it follows that $v_3\ge 2y_1$.  It is convenient to write \eqref{GR_zto0} as derivative of a delta function:
\begin{align}
G^R_\thT(3|1)=-\frac{1}{4}z_3^2z_1^2\d'(\cosh(x_{31})-\cosh(v_3)+\sinh(v_3)z_1).
\end{align}
Now we can use $G^R_\thT(3|1)$ to propagate point $1$ to point $3$. We will use the delta function to eliminate the integration of $z_1$\footnote{We could eliminate the integration of $x_1$ instead, but then the integration of $z_1$ sees a divergence as $y_1\to v_3/2$. This is not a true divergence but requiring more careful treatment of the delta function.}. Thew following trick will be used:
\begin{align}
&\int dz_1\D G^>(2|1)\oar{D}^{v1}G^R_\thT(3|1) \no
=&\int dz_1\D G^>(2|1)\(-\frac{1}{z_1}\oar{\pd}_{z1}\)G^R_\thT(3|1) \no
=&\int dz_1\frac{\D G^>(2|1)}{\sqrt{z_1}}\(-\oar{\pd}_{z1}\)\frac{G^R_\thT(3|1)}{\sqrt{z_1}} \no
=&\int dz_12\pd_{z1}\(\frac{\D G^>(2|1)}{\sqrt{z_1}}\)\frac{G^R_\thT(3|1)}{\sqrt{z_1}}.
\end{align}
We have redistributed the factor $\frac{1}{z_1}$ into $\D G^>(2|1)$ and $G^R_\thT(3|1)$. This is justified because the terms from derivatives on $\frac{1}{\sqrt{z_1}}$ cancel pairwise. We have also used partial integration in the last step because $G^R_\thT(3|1)$ only has finite support on $z_1$.
Now we are ready to put everything together to obtain
\begin{align}\label{trick_rep}
\D G^>(4|3)=&\int dz_1dx_1dz_2dx_2\pd_{z1}\pd_{z2}\(\frac{\D G^>(2|1)}{\sqrt{z_2z_1}}\)(z_2z_1)^{3/2} \no
&\times\d'(\cosh(x_{31})-\G_1)\d'(\cosh(x_{42})-\G_2),
\end{align}
where
\begin{align}
&\G_1=\cosh(v_3)-\sinh(v_3)z_1, \no
&\G_2=\cosh(v_4)-\sinh(v_4)z_2.
\end{align}
Careful elimination of the delta functions in \eqref{trick_rep} leads to
\begin{align}\label{final_rep}
\D\<O(v_4,x_4)O(v_3,x_3)\>&=\int dx_{31}dx_{42}\frac{\pd{z_1}}{\pd\G_1}\frac{\pd{z_2}}{\pd\G_2}\pd_{z1}\pd_{z2}\bigg[\frac{\pd{z_1}}{\pd\G_1}\frac{\pd{z_2}}{\pd\G_2}\pd_{z1}\pd_{z2}\(\frac{i\D G^>(2|1)}{\sqrt{z_2z_1}}\)(z_2z_1)^{3/2}\bigg] \no
&=\frac{1}{\sinh^2v_3\sinh^2v_4}\int dx_{31}dx_{42}\pd_{z1}\pd_{z2}\bigg[\pd_{z1}\pd_{z2}\(\frac{i\D G^>(2|1)}{\sqrt{z_2z_1}}\)(z_2z_1)^{3/2}\bigg],
\end{align}
with $z_1=\frac{\cosh(v_3)-\cosh(x_{31})}{\sinh(v_3)}$ and $z_2=\frac{\cosh(v_4)-\cosh(x_{42})}{\sinh(v_4)}$. The integrations of $x_{31}$ and $x_{42}$ are bounded by $|x_{31}|\le v_3$ and $|x_{42}|\le v_4$ respectively.
By construction, $\D\<O(v_4,x_{43})O(v_3,0)\>$ is the difference of correlation in thermalizing state minus the counterpart in thermal state:
\begin{align}\label{oo_def}
\D\<O(v_4,x_4)O(v_3,x_3)\>=\<O(v_4,x_{43})O(v_3,0)\>-\<O(v_4,x_{43})O(v_3,0)\>_\text{th}.
\end{align}
We will evaluate \eqref{final_rep} in different kinematic regimes and show the results in the next section.

\section{Results in coordinate space}

We will be interested in two classes of correlators: i. equal-time correlator $\D\<O(v,x)O(v,0)\>$; ii. equal-space correlator $\D\<O(v_4,0)O(v_3,0)\>$. The first class measures evolution of spatial correlation, and the second class measures the evolution of temporal correlation. We will evaluate them separately.

\subsection{Equal-time correlator}

We start from the definition \eqref{oo_def}. For equal-time correlator, only the first term evolves with time $v$. The second term is stationary. At initial time of thermalization $v=0$, we expect the first term to reduce to vacuum correlator. According to \eqref{cft_wight}, we obtain
\begin{align}\label{veq0}
\D\<O(v,x)O(v,0)\>\xrightarrow{v\to 0} \frac{2}{\pi}\frac{1}{x^4}-\frac{1}{2\pi}\frac{1}{\(-1+\cosh x\)^2}.
\end{align}
We see that the first term is a power law decay in separation, which indicates long range correlation in vacuum of CFT. The second term features an exponential decay, which characterizes Debye screening, with the screening length set by $1/T\sim 1$ in our unit.
From now, we will now specialize to two limiting cases: $x\gg 1$ and $x\ll 1$. The two cases correspond to the spatial separation much larger/smaller than the spatial screening length.
 
\begin{figure}[t]
\includegraphics[width=0.5\textwidth]{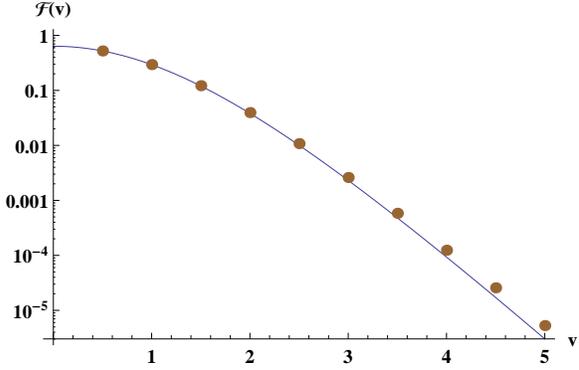}
\caption{\label{fig_f}${\cal F}(v)=x^4\D \<O(v,x)O(v,0)\>$ as a function of $v$ for $x=15$. The analytic fitting function is given by $\frac{18}{\pi}\(\frac{v\coth v-1}{\sinh^2v}\)^2$. It fits very well in a wide range of $v$.}
\end{figure}

\begin{figure}[t]
\includegraphics[width=0.5\textwidth]{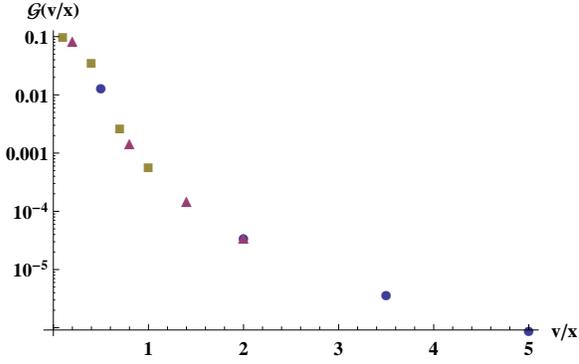}
\caption{\label{fig_g}${\cal G}(v/x)=x^2\D \<O(v,x)O(v,0)\>$ as a function of $v/x$ for $x=1/50$(blue point), $x=1/20$(purple triangle) and $x=1/10$(brown square). The tail is numerically consistent with a power law ${\cal G}\(\frac{v}{x}\)\sim \(\frac{x}{v}\)^4$.}
\end{figure}

When $x\gg 1$, we can ignore the exponential decaying term. $\D\<O(v,x)O(v,0)\>$ simply measures decay of vacuum correlator. For this reason, we parametrize the correlator as $\D\<O(v,x)O(v,0)\>=\frac{1}{x^4}{\cal F}(v)$. The decaying function ${\cal F}(v)$ can be obtained from numerical integration of \eqref{final_rep}. Figure~\ref{fig_f} shows numerical results of ${\cal F}(v)$. We also include analytic approximation of ${\cal F}(v)$. The specific form of the analytic expression will be obtained later.
Next we consider $x\ll 1$. In this case, the leading $\frac{1}{x^4}$ behavior in vacuum and thermal correlators cancel out. \eqref{veq0} reduces to
\begin{align}
\D\<O(v,x)O(v,0)\>\xrightarrow{v\to 0}\frac{1}{3\pi}\frac{1}{x^2}.
\end{align}
We further restrict ourselves to the regime $v\ll 1$. In this regime, we expect the following scaling $\D\<O(v,x)O(v,0)\>=\frac{1}{x^2}{\cal G}\(\frac{v}{x}\)$. In Figure~\ref{fig_g}, we confirm the scaling behavior by showing numerical results of ${\cal G}\(\frac{v}{x}\)$ obtained with different $x$ agree with each other. However, we do not have an analytic expression for the scaling function ${\cal G}\(\frac{v}{x}\)$.

\subsection{Analytic results for spacelike correlator}

It is a good point to present analytic results for spacelike correlator $\<O(v_4,x_{43})O(v_3,0)\>$, which include equal-time correlator discussed in the previous subsection. One limit allowing for analytic treatment is $x_{43}\gg 1$, $x_{43}\gg v_3,\; v_4$. In this case, $x_{21}=x_{43}-x_{42}+x_{31}\simeq x_{43}$ because $|x_{31}|\le v_3$, $|x_{42}|\le v_4$ and $|x_{43}|\gg v_3,v_4$. 
Since $x_{21}\simeq x_{43}>0$, we have $\xi<1$ thus regularization is not needed. We can calculate $\<O(v_4,x_{43})O(v_3,0)\>$ directly. This is possible as long as $x_{43}>v_4+v_3$, which can be viewed as a generalized spacelike condition for correlator $\<O(v_4,x_{43})O(v_3,0)\>$.
To proceed, we note $x_{43}\gg 1>z_1z_2$, therefore we can approximate the invariant distance in AdS
\begin{align}
&\x_\ads=\frac{2z_2z_1}{2z_2z_2+x_{21}^2}\simeq\frac{2z_2z_1}{x_{43}^2}\ll1. 
\end{align}
Note that $G_E(\x)\sim \x^2$ as $\x\to 0$. As a result, $\D G^>(2|1)$ simplifies significantly:
\begin{align}\label{DG_i}
i G^>(2|1)\simeq \frac{\x_\ads^2}{8\pi}\simeq \frac{1}{8\pi}\(\frac{2z_2z_1}{x_{43}^2}\)^2.
\end{align}
With \eqref{DG_i}, we can evaluate \eqref{final_rep} analytically to obtain
\begin{align}\label{DO_i}
\<O(v_4,x_4)O(v_3,x_3)\>\simeq \frac{18}{\pi x_{43}^4}\frac{v_3\coth(v_3)-1}{\sinh^2(v_3)}\frac{v_4\coth(v_4)-1}{\sinh^2(v_4)}.
\end{align}
Note that \eqref{DO_i} is valid for arbitrary $v_3,v_4$ as long as $x_{43}\gg v_3,v_4$. \eqref{DO_i} gives the fitting function in Fig.~\ref{fig_f} upon setting $v_3=v_4=v$.
It gives for $v_3,v_4\ll 1$
\begin{align}\label{i_small_v}
\<O(v_4,x_4)O(v_3,x_3)\>\simeq \frac{2}{\pi x_{43}^4},
\end{align}
which reproduces Eq.~(5.7) of \cite{Aparicio:2011zy}. 
On the other hand, for $v_3,v_4\gg 1$
\begin{align}\label{i_large_v}
\<O(v_4,x_4)O(v_3,x_3)\>\simeq \frac{18}{\pi x_{43}^4}v_3v_4e^{-2v_3-2v_4},
\end{align}
while the geodesic approximation in \cite{Aparicio:2011zy} gives
\begin{align}\label{geo_approx}
\<O(v_4,x_4)O(v_3,x_3)\>\sim \frac{1}{x_{43}^4}e^{-2v_3-2v_4}.
\end{align}
We note that the geodesic approximation misses the enhancement factor $v_3v_4$ in spacelike correlator. It is not difficult to understand the reason from gravity point of view: since geodesic approximation only knows about the bulk geometry lying between boundary insertion times $v_3$ and $v_4$, while the enhancement factor results from the history of the bulk geometry (from the time of energy injection $v=0$ to the times of measurement at $v_3$ and $v_4$).

It is informative to compare our results with general results obtained by CC \cite{Calabrese:2006rx}. In the latter case, the correlator $\<O(v_4,x)O(v_3,0)\>$ in a thermalizing state is given by
\begin{align}\label{cc_result}
\<O(v_4,x_{43})O(v_3,0)\>\sim
\left\{\begin{array}{l@{\quad}l}
e^{-2\pi(v_3+v_4)/2\t_0}& x_{43}>v_3+v_4\\
e^{-2\pi x_{43}/2\t_0}& |v_4-v_3|<x_{43}<v_3+v_4\\
e^{-2\pi|v_4-v_3|/2\t_0}& x_{43}<|v_4-v_3|
\end{array}
\right.
\end{align}
$\t_0$ is proportional to inverse temperature adapted to our model. Comparing \eqref{cc_result} in the long time (thermal) limit with \eqref{cft_wight}, we identify $2\t_0=\pi$. However, \eqref{cc_result} does not contain the power law factor $1/x_{43}^4$ present in \eqref{i_large_v}. As already pointed out in \cite{Aparicio:2011zy}, it is because the initial state in the formulation of CC has finite correlation length, while the power law is reminiscent of long range correlation present in the initial state modeled by Vaidya background. We argue now the time factor $v_3v_4$ missed in geodesic approximation is also due to the long range correlation in the initial state: as $\<\D O(v_4,x_4)O(v_3,x_3)\>$ receives contribution from regions in the backward lightcone, factors of $v_3$ and $v_4$ come from the distance of the correlated regions that propagate initial state correlation to point $x_3$ and $x_4$ respectively. 

We note that in case of CC \eqref{cc_result}, a spacelike correlator $\<O(v_4,x_{43})O(v_3,0)\>$ changes from exponential decaying form $e^{-2(v_3+v_4)}$ to thermal form $e^{-2x_{43}}$ when $v_3+v_4$ exceeds $x_{43}$, upto correction of order inverse temperature. 
If we define thermalization time to be the largest possible value of $v_3$ or $v_4$ across which the correlator $\<O(v_4,x_{43})O(v_3,0)\>$ appears thermal, \eqref{cc_result} implies that points separated by distance $x_{43}$ takes a time $x_{43}+O(1)$ to thermalize \footnote{This is realized when $v_4\to x_{43}$ and $v_3\to0$ for example.}. When the initial state has long range correlation, the spacelike correlator $\<\D O(v_4,x_{43})O(v_3,0)\>$ is modified by power law factor $v_3v_4/x_{43}^4$. This seems to imply that the thermalization time could be modified to $x_{43}+O(\ln x_{43})$. We will show below that it is not the case.

\begin{figure}[t]
\includegraphics[width=0.5\textwidth]{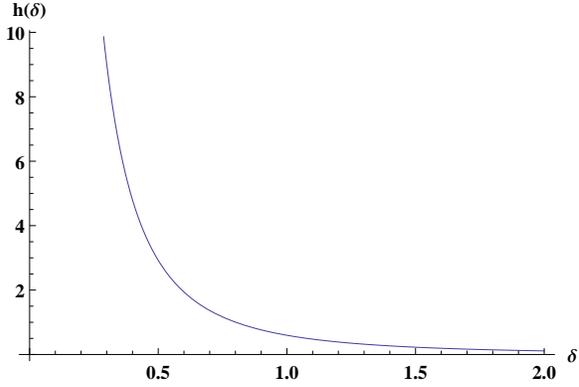}
\caption{\label{fig_delta}$h(\d)\equiv \frac{1}{\d^2}-\frac{1}{\d}-e^{\d}Ei(-\d)$ as a function of $\d$. It is a monotonously decreasing function of $\d$. As $\d\to0$, the two insertion points $(v_4,x_{43}=v_4+\d)$ and $(v_3,0)$($v_3\ll 1$) approach the lightcone. The dropping of $h(\d)$ corresponds to the thermalization of correlator, which occurs on a time scale of $O(1)$.}
\end{figure}

It is desirable to find analytic results for correlator near the ``lightcone'': $v_3+v_4\simeq x_{43}$. It turns out to be possible in the following regime $x_{43}\gg 1$ and $x_{43}-v_4\equiv \d\sim O(1) \gg v_3$. We consider $\d>0$, which satisfies generalized spacelike condition $x_{43}-v_4-v_3>0$, thus regularization is not needed. We will evaluate $\<O(v_4,x_{43})O(v_3,0)\>$ directly and compare to the thermal correlator $\<O(0,x_{43})O(0,0)\>_\text{th}$. In the regime we work in, $x_{21}=x_{43}-x_{42}+x_{31}>x_{43}-v_4-v_3\simeq \d$. Furthermore $z_1<\frac{\cosh v_3-1}{\sinh v_3}=\tanh\frac{v_3}{2}\ll \d$ and $z_2<1$, thus $x_{21}\gg z_1z_2$. Combining the above, we can approximate
\begin{align}
\x_{\ads}\simeq\frac{2z_2z_1}{x_{43}-x_{42}}.
\end{align}
The integration of $x_{31}$ and $x_{42}$ factorizes. The integration of $x_{31}$ and $x_{42}$ can be done separately as follows:
\begin{align}\label{x_int}
&\int_{-v_3}^{v_3} dx_{31}\frac{z_1}{\sinh^2v_3}=\frac{2(v_3\coth v_3-1)}{\sinh^2v_3}, \no
&\int_{-v_4}^{v_4} dx_{42}\frac{z_2}{\sinh^2v_4}\frac{1}{(x_{43}-x_{42})^4}=\frac{1}{\sinh^3v_4}\big[\frac{\cosh v_4-\cosh x_{42}}{3(x_{43}-x_{42})^3}+\frac{\sinh x_{42}}{6(x_{43}-x_{42})^2}-\frac{\cosh x_{42}}{6(x_{43}-x_{42})} \no
&-\frac{1}{12}\(e^{x_{43}}Ei(x_{42}-x_{43})-e^{-x_{43}}Ei(x_{43}-x_{42})\)\big]|_{x_{42}=-v_4}^{v_4}.
\end{align}
Working in the limit $v_4\gg 1,\; v_3\ll 1$ of \eqref{x_int}, we find the $x_{31}$ integral approaches a constant, and the $x_{42}$ integral asymptotes to
\begin{align}\label{delta}
\frac{e^{-2v_4}}{12}\(\frac{1}{\d^2}-\frac{1}{\d}-e^{\d}Ei(-\d)\).
\end{align}
The correlator $\<O(v_4,x_{43})O(v_3,0)\>$ given by the product of two integrals in \eqref{x_int} (upto overall numerical factor) does not have enhancement factor $v_4$ close to the lightcone. Comparing \eqref{delta} with thermal correlator $e^{-2x_{43}}$, we conclude that the thermalization time in this case is given by $v_4+O(1)\simeq x_{43}+O(1)$, thus free from $\ln x_{43}$ correction.
For completeness, we also plot the function in the bracket of \eqref{delta} in Figure~\ref{fig_delta}, which is a monotonously decreasing function of $\d$. 

We have also performed numerical studies of equal-time correlator $\<O(v,x)O(v,0)\>$ for large $v$ and $x$ near the generalized lightcone $x=2v$. Defining the point at which equal-time correlator drops to twice the thermal correlator as the thermalization time, we find the thermalization time is still given by $v\simeq x+O(1)$, free from correction of order $\ln x$.

\subsection{equal-space correlator}

\begin{figure}[t]
\includegraphics[width=0.5\textwidth]{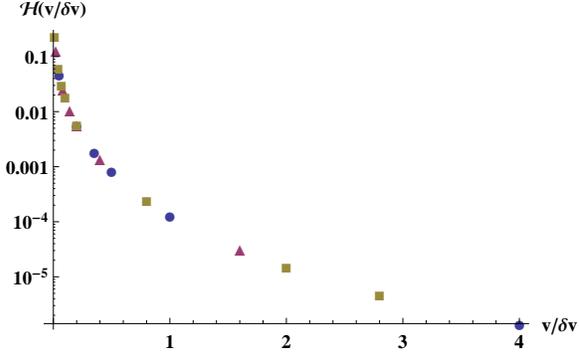}
\caption{\label{fig_h}${\cal H}\(v/\d v\)=\d v^2\D \<O(v+\d v,0)O(v,0)\>$ as a function of $v/\d v$ for $\d v=1/50$(blue point), $\d v=1/20$(purple triangle) and $\d v=1/10$(brown square) fall onto the same scaling curve.}
\end{figure}

Now we look at equal-space correlator $\D\<O(v_4,0)O(v_3,0)\>$ with $v_4\equiv v+\d v$ and $v_3\equiv v$. Parallel to the equal-time case, we study the regimes $\d v\ll 1$ and $\d v\gg 1$.
Similar to equal-time correlator, we might expect equal-space correlator $\D\<O(v+\d v,0)O(v,0)\>$ to reduce to the difference between vacuum and thermal correlator as $v\to 0$, which is
\begin{align}\label{veq0_es}
\D\<O(v+\d v,0)O(v,0)\>\xrightarrow{v\to 0} \frac{2}{\pi}\frac{1}{\d v^4}-\frac{1}{2\pi}\frac{1}{\(-1+\cosh\d v\)^2}.
\end{align}
In the regime $\d v\ll 1$, that is when the temporal separation much less than the screening length, \eqref{veq0_es} reduces to $\frac{1}{3\pi}\frac{1}{\d v^2}$. This motivates the following scaling behavior for $\D\<O(v+\d v,0)O(v,0)\>=\frac{1}{\d v^2}{\cal H}\(\frac{v}{\d v}\)$ when $\d v\ll 1,\; v\ll 1$. Indeed, we can confirm the scaling behavior from numerical results. Figure~\ref{fig_h} shows the scaling function ${\cal H}\(\frac{v}{\d v}\)$ for different $\d v$.
However, the expectation \eqref{veq0_es} turns out to be incorrect. The expectation would predict ${\cal H}\(\frac{v}{\d v}\)$ approaches the constant $\frac{1}{3\pi}$, while the numerical results blow up as $\frac{v}{\d v}\to 0$.
Interestingly, if we consider the limit $\d v\to 0$, the correlator $\D\<O(v+\d v,0)O(v,0)\>$ always gives a finite value. This shows a non-commutativity between the limit $v\to 0$ and $\d v\to 0$.

\begin{figure}[t]
\includegraphics[width=0.5\textwidth]{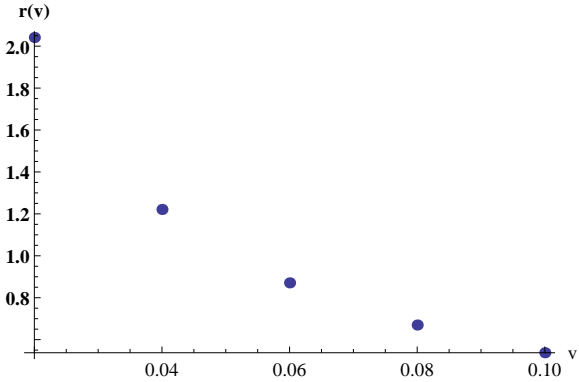}
\caption{\label{fig_r}$r(v)=\D\<O(v+\d v,0)O(v,0)\>/\<O(\d v,0)O(0,0)\>_\thT$ as a function of $v$ for $\d v=10$. It is a monotonously decreasing function of order one, indicating a thermalization time of order inverse temperature for large temporal interval.}
\end{figure}

Now we study the regime $\d v\gg 1$. We would like to find the thermalization time for temporal interval $\d v\gg 1$. For equal-space correlator, CC results \eqref{cc_result} implies a thermalization time of order inverse temperature, independent of $\d v$. We define thermalization time to be $v$ at which $\D\<O(v+\d v,0)O(v,0)\>$ drops to the thermal correlator $\<O(\d v,0)O(0,0)\>_\thT$. We perform numerical studies and find $\D\<O(v+\d v,0)O(v,0)\>\sim e^{-2\d v}$ for $\d v\gg 1$ and $v=O(1)$. On the other hand, $\<O(\d v,0)O(0,0)\>_\thT\sim e^{-2\d v}$. Figure~\ref{fig_r} includes a plot of $r(v)\equiv \D\<O(v+\d v,0)O(v,0)\>/\<O(\d v,0)O(0,0)\>_\thT$.

Summarizing this section, we have found analytic expressions of equal-time correlator and spacelike correlator near the lightcone. While the expressions deviate from CC and holographic results obtained with geodesic approximation, the thermalization time is not significantly modified due to the presence of long range correlation in the initial state: a thermalization time $x_{43}+O(1/T)$ for spacelike correlator $\D\<O(v_4,x_{43})O(v_3,0)\>$ and a thermalization time $O(1/T)$ for equal-space correlator $\D\<O(v+\d v)O(v,0)\>$.

\section{Results for spatially integrated correlator}
In this section, we will work in (spatial) momentum space, but still use temporal coordinates. In particular, we will focus on the spatially integrated correlator, i.e. the mode with $k=0$. Fourier transform of \eqref{final_rep} gives us
\begin{align}\label{k0_rep}
&\int dx_{43}\D\<O(v_4,x_4)O(v_3,x_3)\>=\int dx_{31}dx_{42}\frac{\pd{z_1}}{\pd\G_1}\frac{\pd{z_2}}{\pd\G_2}\pd_{z1}\pd_{z2}\bigg[\frac{\pd{z_1}}{\pd\G_1}\frac{\pd{z_2}}{\pd\G_2}\times \no
&\pd_{z1}\pd_{z2}\(\int dx_{21}\frac{i\D G^>(2|1)}{\sqrt{z_2z_1}}\)(z_2z_1)^{3/2}\bigg],
\end{align}
The spatially integrated initial value is given by
\begin{align}
\int dx_{21}i\D G^>(2|1)=\int dx_{21}i\(G^>_0(2|1)-G^>_\thT(2|1)\).
\end{align}
The integration of $G^>_0(2|1)$ is easily done, with the following result
\begin{align}\label{G0_lc}
\int dx_{21}iG^>_0(2|1)=\frac{\sqrt{b}}{2\pi}Q_{1/2}(1-\frac{\e}{b}),
\end{align}
where $b=z_2z_1$ and $\e=(v_2-v_1)(z_2-z_1)$.
The integration of $iG^>_\thT(2|1)$ requires some effort
\begin{align}\label{Gth_s}
&\frac{1}{4\pi}\int_{-\infty}^{+\infty} dx_{21}\(\frac{1}{\sqrt{1-\x_\btz^2}}-1\) \no
=&\frac{1}{2\pi}\int_1^\infty \frac{ds}{\sqrt{s^2-1}}\(\frac{s-a}{\sqrt{(s-a+b)(s-a-b)}}-1\),
\end{align}
where $a=\sqrt{1-z_2^2}\sqrt{1-z_1^2}\(\cosh(v_2-v_1+y_2-y_1)\)$. In the limit $v_2,v_1\to0$ relevant for our initial condition, $a=1-b+\e$.
In calculating the integral in \eqref{Gth_s}, we note that separate integrations of two terms in the bracket both diverge, but the divergences cancel out in the their difference yielding a finite result. We will calculate \eqref{Gth_s} from the regulated integral
\begin{align}\label{s_int}
&\int_1^\L \frac{ds}{\sqrt{s^2-1}}\(\frac{s-a+b}{s-a-b}\)^{1/2}+
\int_1^\L \frac{ds}{\sqrt{s^2-1}}\(\frac{s-a-b}{s-a+b}\)^{1/2}-
2\int_1^\L \frac{ds}{\sqrt{s^2-1}}.
\end{align}
The limit $\L\to\infty$, $\e\to0$ of the above integral can be obtained analytically. We will only show the final result and collect technical details in Appendix~\ref{app}.
\begin{align}\label{Gth_lc}
\int dx_{21}iG^>_\thT(2|1)=\frac{\(-4\ln(1+\sqrt{b})-\sqrt{b}\(2\ln(1+\frac{1}{\sqrt{b}})+\ln\frac{\e}{32}\)\)}{4\pi}.
\end{align}
We see that although \eqref{G0_lc} and \eqref{Gth_lc} contain separate logarithmic divergences in $\e$ (lightcone singularities), the divergences cancel out in their difference:
\begin{align}\label{DG_lc}
\int dx_{21}i(G_0^>(2|1)-G_\thT^>(2|1))=\frac{\(-2\sqrt{b}+(2+\sqrt{b})\ln(1+\sqrt{b})\)}{2\pi}.
\end{align}
Plugging \eqref{DG_lc} into \eqref{k0_rep}, we obtain the following simple representation
\begin{align}\label{ppDG}
&\int dx_{43}\D\<O(v_4,x_4)O(v_3,x_3)\>=\frac{3}{16\pi\sinh^2(v_3)\sinh^2(v_4)}
\int dx_{31}dx_{42}\frac{\sqrt{z_2z_1}}{(1+\sqrt{z_2z_1})^4},
\end{align}
with $z_1=\frac{\cosh(v_3)-\cosh(x_{31})}{\sinh(v_3)}$ and $z_2=\frac{\cosh(v_4)-\cosh(x_{42})}{\sinh(v_4)}$. We will consider the regime $v_3\ll 1$ and $v_4$ arbitrary. This regime allows us to do the integral analytically. Note that $z_1\le \frac{\cosh(v_3)-1}{\sinh(v_3)}\le v_3\ll1$ and $z_2\le 1$. We can then drop $\sqrt{z_2z_1}$ in the denominator and the integral can be expressed in terms of Elliptic integrals
\begin{align}\label{DO_k0}
&\int dx_{43}\D\<O(v_4,x_4)O(v_3,x_3)\>\simeq \frac{3\sqrt{2}}{16\sqrt{v_3}}\frac{\sqrt{\coth(v_4/2)}}{\sinh^2(v_4)}\(K(\tanh(v_4/2))-E(\tanh(v_4/2))\).
\end{align}
We consider the early time and late time regime of \eqref{DO_k0}. At early time $v_4\ll 1$, we obtain
\begin{align}\label{k0_small_v}
\int dx_{43}\D\<O(v_4,x_4)O(v_3,x_3)\>\simeq \frac{3\pi}{128\sqrt{v_3v_4}}.
\end{align}
The dependence $\frac{1}{\sqrt{v_3v_4}}$ can be understood in the following way: We claim that the spatial integration receives dominant contribution from the domain with $x_{43}\ll 1$, i.e. short distance. 
This is most easily seen in equal-time correlator $\D\<O(v,x)O(v,0)\>$. For $v,\; x\ll 1$, the equal-time correlator is given by
\begin{align}
\D\<O(v,x)O(v,0)\>=\frac{1}{x^2}{\cal G}\(\frac{v}{x}\).
\end{align}
Integrating the above over $x$, we obtain $\sim\frac{1}{v}$. Numerically integrating the scaling function ${\cal G}\(\frac{v}{x}\)$, we can confirm the numerical factor in \eqref{k0_small_v}. Therefore, the spatially integrated equal-time correlator in far from equilibrium regime receives dominant contribution from short distance physics, in contrast to the equilibrium intuition that small momentum is equivalent to long distance physics. With more sophisticated analysis, we could show that the conclusion remains true for more generic correlator $\D\<O(v_4,x_{43})O(v_3,0)\>$.
At late time $v_4\gg 1$, we obtain
\begin{align}
\int dx_{43}\D\<O(v_4,x_4)O(v_3,x_3)\simeq \frac{3\sqrt{2}}{32\sqrt{v_3}}v_4e^{-2v_4}.
\end{align}

\subsection{Out-of-equilibrium emission spectrum}

\begin{figure}[t]
\includegraphics[width=0.5\textwidth]{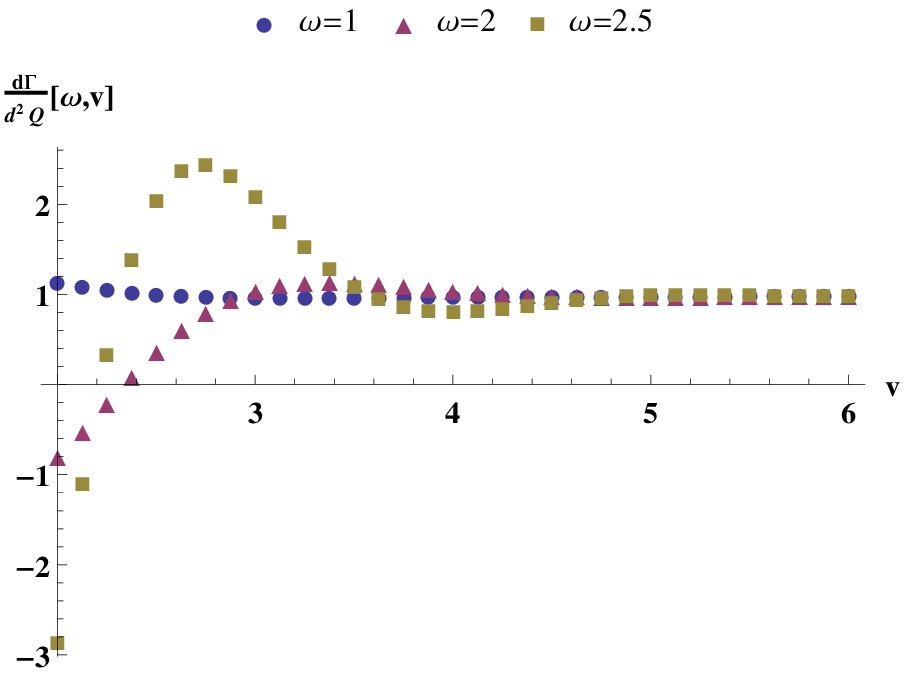}
\includegraphics[width=0.5\textwidth]{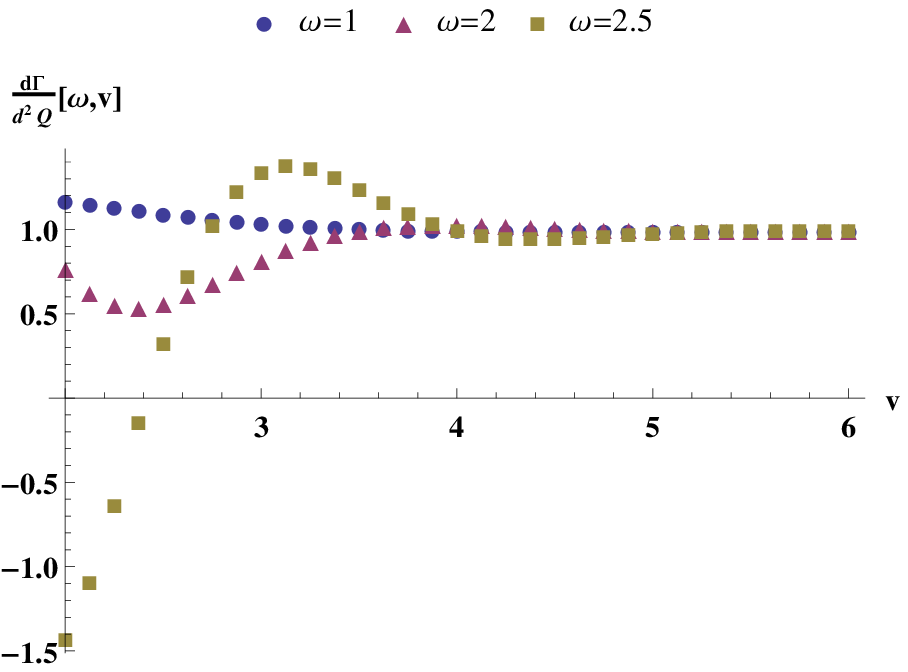}
\caption{\label{fig_Gm}Emission rates $\frac{d\G}{d^2Q}\(\o,v\)$ as a function of $v$ in the thermalizing state (left) and in an instantaneous thermalization scenario (right), see \eqref{diff_rep}-\eqref{inst_rate} for corresponding definitions. Both rates are in unit of thermal emission rate $\frac{d\G_\thT}{d^2Q}\(\o\)$, defined in \eqref{th_rate}. The symbols represent $\o=1$ (blue point), $\o=2$ (purple triangle) and $\o=2.5$ (brown square). All the rates approach thermal limit at late time. We observe a hierarchy among spectra with different frequencies, with lower frequency mode tends to thermal spectrum faster. The high frequency mode shows more oscillations in relaxing to thermal spectrum. A possible cause for the negative rate at early time is given in the text.}
\end{figure}

As an application, we calculate a physical observable: emission spectrum of particles weakly coupled to operator $O$. We can draw analogy with dilepton emission: we can regard $O$ as current, and radiated particles as dilepton. The coupling constant $g_O$ between radiated particle and $O$ is small like the electromagnetic coupling $e$. With an abuse of terminology, we will refer to the radiated particle as dilepton and the field created by $O$ as photon. In the absence of translational invariance in time, we use the following operational definition for local emission rate of dilepton: as dilepton is being radiated continuously in the thermalization process, we define the differential yield at time $v$ as the emission rate. We formulate the differential rate as follows: the transition amplitude from an initial state $|i\>$ to a final state $|f\>$ with photon is given by
\begin{align}
S_{fi}=g_O\int d^2Xe^{iQ X}\<f|O(X)|i\>.
\end{align}
The total yield of dilepton is given by
\begin{align}\label{tot_S}
\sum_{f}|S_{fi}|^2&=g_O^2\int d^2Xd^2Ye^{iQ(X-Y)}\sum_f\<i|O(Y)|f\>\<f|O(X)|i\> \no
&=g_O^2\int d^2Xd^2Ye^{iQ(X-Y)}\<i|O(Y)O(X)|i\>.
\end{align} 
We count the yield through time $v$, thus the integration of $X^0$ and $Y^0$ is taken from $-\infty$ to $v$. Using translational invariance in spatial direction, we can simplify the total rate as
\begin{align}\label{tot_St}
\sum_{f}|S_{fi}|^2=g_O^2\int_{-\infty}^v dX^0\int_{-\infty}^vdY^0V\int d(X^1-Y^1)e^{iQ(X-Y)}\<i|O(Y)O(X)|i\>,
\end{align}
where $V$ is the one dimensional volume. We have used spatial translation invariance in \eqref{tot_St}.
Now we identify \eqref{tot_St} with spacetime integral of differential emission rate
\begin{align}
V\int_{-\infty}^v dt\frac{d\G(t)}{d^2Q}=g_O^2\int_{-\infty}^v dX^0\int_{-\infty}^vdY^0V\int d(X^1-Y^1)e^{iQ(X-Y)}\<i|O(Y)O(X)|i\>.
\end{align}
Canceling the volume factor and taking the derivative with respect to $v$, we obtain the following representation of differential rate
\begin{align}\label{diff_rep}
\frac{d\G(\o, v)}{d^2Q}&=g_O^2\big[\int_{-\infty}^v dX^0e^{i\o(X^0-v)}\vert_{Y^0=v}+\int_{-\infty}^v dY^0e^{i\o(v-Y^0)}\vert_{X^0=v}\big] \no
&\times \int d(X^1-Y^1)e^{-ik(X^1-Y^1)}\<i|O(Y)O(X)|i\>.
\end{align}
For $k=0$, the regularized version of the quantity $\int d(X^1-Y^1)e^{-ik(X^1-Y^1)}\<i|O(Y)O(X)|i\>$, has already been calculated in \eqref{ppDG}. To obtain the full result, we add back the thermal part:
\begin{align}\label{ppDG_full}
\int d(X^1-Y^1)\<i|O(Y)O(X)|i\>=\int dx_{43}\(\D\<O(v_4,x_{43})O(v_3,0)\>+\<O(v_4,x_{43})O(v_3,0)\>_\text{th}\).
\end{align}
The thermal part is given by
\begin{align}\label{ppDG_th}
\int dx_{43}\<O(v_4,x_{43})O(v_3,0)\>_\text{th}&=\int dx_{43}\frac{1}{2\pi}\frac{1}{\(-\cosh(v_4-v_3-i\e)+\cosh x_{43}\)^2} \no
&=\frac{1}{2\pi}\frac{(v_4-v_3)\coth(v_4-v_3)-1}{\sinh^2(v_4-v_3)}.
\end{align}
From \eqref{ppDG} and \eqref{ppDG_th}, we can see that $\int d(X^1-Y^1)\<i|O(Y)O(X)|i\>$ is invariant under exchange of $X$ and $Y$. It is not difficult to show that the exchange symmetry leads to the spectrum being an even function of $\o$.
Note that \eqref{ppDG} is valid only after $v>0$, therefore the integration of $dX^0$ and $dY^0$ start from $t=0$ through $t=v$. Before $v=0$, the state is vacuum state, which does not radiate any dilepton. To compare the emission rate with different frequencies, we normalize the rate by dividing the thermal rate, which is given by the Fourier transform of \eqref{ppDG_th}
\begin{align}\label{th_rate}
\frac{d\G(\o)_\thT}{d^2Q}(\o)=g_O^2\int dt e^{-i\o t}\frac{1}{2\pi}\frac{t\coth t-1}{\sinh^2t}.
\end{align}
As a separate reference, we also calculate the emission rate, assuming instantaneous thermalization of the state at $v=0$. The emission rate in this case is given by
\begin{align}\label{inst_rate}
\frac{d\G_\text{inst}(\o, v)}{d^2Q}&=g_O^2\big[\int_{0}^v dX^0e^{i\o(X^0-v)}\vert_{Y^0=v}+\int_{0}^v dY^0e^{i\o(v-Y^0)}\vert_{X^0=v}\big] \no
&\times \int d(X^1-Y^1)\<i|O(Y)O(X)|i\>_\thT.
\end{align}
We present our results for $\o=1/2$, $\o=1$ and $\o=4$ in Figure~\ref{fig_Gm}. We see both $d\G(\o,v)/d^2Q$ and $d\G_\text{inst}(\o,v)/d^2Q$ approach the thermal rate at large time. The deviation of the former comes from out-of-equilibrium effect and missing radiation before $v=0$. The deviation of the latter includes only missing radiation before $v=0$. We do observe the rates become negative at early time. We believe this is an artifact of our definition: strictly speaking, we need to know the past and future of radiation in order to define plane wave spectrum, while we only know the history upto our measurement point at $v$. 
With the caveat in mind, we do observe interesting hierarchy in frequency: low frequency mode tends to appear thermal faster than high frequency mode. Previous studies have shown that short distance physics tends to thermalize faster than long distance physics. Our results can be viewed as a complementary picture to this, although in a counter intuitive way. We also note that high frequency mode shows more oscillations in relaxing to thermal spectrum.

\section{Acknowledgments}
The author is grateful to E.~Shuryak and D.~Teaney for insightful discussions, which significantly improve this work. He would like to thank S.~Schlichting, S.~Stricker and H.-U.~Yee for useful discussions. He also thank the Institute of Nuclear Theory for hospitality at the workshop ``Equilibration Mechanisms in Weakly and Strongly Coupled Quantum Field Theory'' in the completion of this work. This work is in part supported by RIKEN Foreign Postdoctoral Researcher Program and Junior Faculty's Fund of Sun Yat-Sen University.

\appendix

\section{Evaluation of \eqref{s_int}}\label{app}

We reproduce \eqref{s_int} below for easy reference
\begin{align}\label{s_int2}
&\int_1^\L \frac{ds}{\sqrt{s^2-1}}\(\frac{s-a+b}{s-a-b}\)^{1/2}+
\int_1^\L \frac{ds}{\sqrt{s^2-1}}\(\frac{s-a-b}{s-a+b}\)^{1/2}-
2\int_1^\L \frac{ds}{\sqrt{s^2-1}}.
\end{align}
We are interested in the $\L\to\infty$, $\e=1-a-b\to0$ limit of \eqref{s_int2}. The limit $\e\to0$ of the second integral can be taken directly, after which the integral can be expressed in terms of elementary function. In taking the limit $\L\to\infty$, we only need to keep upto constant terms
\begin{align}\label{second}
\lim_{\e\to0,\L\to\infty}\int_1^\L \frac{ds}{\sqrt{s^2-1}}\(\frac{s-a-b}{s-a+b}\)^{1/2}&=\lim_{\L\to\infty}\int_1^\L ds\(\frac{1}{(s+1)(s-1+2b)}\)^{1/2} \no
&=2\ln 2-2\ln(\sqrt{2}+\sqrt{2b})+\ln\L+O(\L^{-1}).
\end{align}
The third term is done in a similar way:
\begin{align}\label{third}
\lim_{\L\to\infty}\int_1^\L\frac{ds}{\sqrt{s^2-1}}=\ln 2+\ln\L+O(\L^{-1}).
\end{align}
The evaluation of the first term needs some effort. The first term can be expressed in terms of elliptic integrals. The formal expression is not very helpful in obtaining asymptotics. We will instead use the following representation
\begin{align}\label{first_rep}
&\int_1^\L\frac{ds}{\sqrt{s^2-1}}\(\frac{s-a+b}{s-a-b}\)^{1/2} \no
=&\frac{2}{\sqrt{(1-a+b)(1+a+b)}}\big[(1-a-b)\int_0^{\sin\n}\frac{dx}{\(1-\frac{2}{1+a+b}x^2\)\sqrt{(1-x^2)(1-q^2x^2)}} \no
&+2b\int_0^{\sin\n}\frac{dx}{\sqrt{(1-x^2)(1-q^2x^2)}}\big], \no
&\text{where} \no
&\sin\n=\(\frac{(1+a+b)(\L-1)}{2(\L-a-b)}\)^{1/2},\quad q^2=\frac{4b}{(1-a+b)(1+a+b)}.
\end{align}
We first look at the second integral in \eqref{first_rep}. Expanding the denominator of the integrand, we obtain
\begin{align}
\sqrt{(1-x^2)(1-q^2x^2)}=\sqrt{\l\(\l-\frac{1}{2}+\frac{1}{2b}\)}\e+O(\e^2),
\end{align}
where $x^2=1-\e\l$.
The upper bound of $x$, in the limit $\e\to0$ is given by
\begin{align}
\sin\n=1-\frac{\e}{4}\frac{\L+1}{\L-1}+O(\e^2),
\end{align}
which translates to the lower bound of $\l$: $\l\ge \frac{1}{2}\frac{\L+1}{\L-1}$. Assuming that the integral upto constant terms arise from the region $\l\sim O(1)$ or $1-x\sim O(\e)$, we obtain the result for the integral
\begin{align}\label{small_ls}
\frac{\e}{2}\int_{\frac{1}{2}\frac{\L+1}{\L-1}}^{\l_s}d\l \frac{1}{\sqrt{\l\(\l-\frac{1}{2}+\frac{1}{2b}\)}\e}=2\ln2-\ln\(\sqrt{2}+\sqrt{\frac{2}{b}}\)+\frac{1}{2}\ln\l_s,
\end{align}
where $\l_s\sim O(1)$ is cutoff of the integral. We note that there is a logarithmic divergence in $\l_s$, which means that there must be contribution from region $1-x\gg O(\e)$ to cancel the divergence. We evaluate the other contribution below
\begin{align}\label{large_ls}
\int_0^{\sqrt{1-\e \l_s}}\frac{dx}{\sqrt{(1-x^2)(1-q^2x^2)}}=\frac{1}{2}\(-\ln\frac{\e}{4}-\ln\l_s\)+O(\l_s^{-1}).
\end{align}
The logarithmic divergences indeed cancel upon adding \eqref{small_ls} and \eqref{large_ls}.
The term $O(\l_s^{-1})$ can be ignored when we take $1\ll \l_s\ll O(1/\e)$. There is also a $\ln\e$ divergence term, which can be traced back to lightcone singularity integrated over spatial coordinate. This term will be canceled by zero temperature counterpart.
The evaluation of the other integral follows similar procedure. We expand the denominator of the integrand as
\begin{align}
\(1-\frac{2}{1+a+b}x^2\)\sqrt{(1-x^2)(1-q^2x^2)}=\(\l-\frac{1}{2}\)\sqrt{\l}\sqrt{\l-\frac{1}{2}+\frac{1}{2b}}\e^2+O(\e^3).
\end{align}
The integration from region $1-x\sim O(\e)$ gives
\begin{align}\label{one_ls}
\lim_{\L\to\infty}\frac{\e}{2}\int_{\frac{1}{2}\frac{\L+1}{\L-1}}^{\l_s}\frac{1}{\(\l-\frac{1}{2}\)\sqrt{\l}\sqrt{\l-\frac{1}{2}+\frac{1}{2b}}\e^2}=\frac{-2\sqrt{b}\tanh^{-1}(\sqrt{b})+\sqrt{b}\(\ln 2-\ln(1-b)+\ln\L\)}{\e}.
\end{align}
We do not see a dependence on the cutoff $\l_s$, suggesting that we can take $\l_s\to\infty$ safely.
Adding \eqref{small_ls}, \eqref{large_ls} and \eqref{one_ls}, we obtain the final result for \eqref{first_rep} in the limit $\e\to0$, $\L\to\infty$:
\begin{align}\label{first_final}
-2\tanh^{-1}\sqrt{b}+6\sqrt{b}\ln2-2\sqrt{b}\ln\(\sqrt{2}+\sqrt{\frac{2}{b}}\)-\sqrt{b}\ln\e-\ln\frac{1-b}{2\L}.
\end{align} 
As remarked before, the $\ln\L$ term will be canceled by \eqref{second} and \eqref{third} and the $\ln\e$ term will be canceled by zero temperature counterpart.

\bibliographystyle{unsrt}
\bibliography{SYSU}

\begin{thebibliography}{10}

\bibitem{Bhattacharyya:2009uu}
Sayantani Bhattacharyya and Shiraz Minwalla.
\newblock {Weak Field Black Hole Formation in Asymptotically AdS Spacetimes}.
\newblock {\em JHEP}, 09:034, 2009.

\bibitem{Grumiller:2008va}
Daniel Grumiller and Paul Romatschke.
\newblock {On the collision of two shock waves in AdS(5)}.
\newblock {\em JHEP}, 08:027, 2008.

\bibitem{Chesler:2015wra}
Paul~M. Chesler and Laurence~G. Yaffe.
\newblock {Holography and off-center collisions of localized shock waves}.
\newblock 2015.

\bibitem{Chesler:2013lia}
Paul~M. Chesler and Laurence~G. Yaffe.
\newblock {Numerical solution of gravitational dynamics in asymptotically
  anti-de Sitter spacetimes}.
\newblock {\em JHEP}, 07:086, 2014.

\bibitem{Chesler:2010bi}
Paul~M. Chesler and Laurence~G. Yaffe.
\newblock {Holography and colliding gravitational shock waves in asymptotically
  AdS$_5$ spacetime}.
\newblock {\em Phys. Rev. Lett.}, 106:021601, 2011.

\bibitem{Chesler:2009cy}
Paul~M. Chesler and Laurence~G. Yaffe.
\newblock {Boost invariant flow, black hole formation, and far-from-equilibrium
  dynamics in N = 4 supersymmetric Yang-Mills theory}.
\newblock {\em Phys. Rev.}, D82:026006, 2010.

\bibitem{Chesler:2008hg}
Paul~M. Chesler and Laurence~G. Yaffe.
\newblock {Horizon formation and far-from-equilibrium isotropization in
  supersymmetric Yang-Mills plasma}.
\newblock {\em Phys. Rev. Lett.}, 102:211601, 2009.

\bibitem{Bantilan:2012vu}
Hans Bantilan, Frans Pretorius, and Steven~S. Gubser.
\newblock {Simulation of Asymptotically AdS5 Spacetimes with a Generalized
  Harmonic Evolution Scheme}.
\newblock {\em Phys. Rev.}, D85:084038, 2012.

\bibitem{Beuf:2009cx}
Guillaume Beuf, Michal~P. Heller, Romuald~A. Janik, and Robi Peschanski.
\newblock {Boost-invariant early time dynamics from AdS/CFT}.
\newblock {\em JHEP}, 10:043, 2009.

\bibitem{Heller:2011ju}
Michal~P. Heller, Romuald~A. Janik, and Przemyslaw Witaszczyk.
\newblock {The characteristics of thermalization of boost-invariant plasma from
  holography}.
\newblock {\em Phys. Rev. Lett.}, 108:201602, 2012.

\bibitem{vanderSchee:2013pia}
Wilke van~der Schee, Paul Romatschke, and Scott Pratt.
\newblock {Fully Dynamical Simulation of Central Nuclear Collisions}.
\newblock {\em Phys. Rev. Lett.}, 111(22):222302, 2013.

\bibitem{Casalderrey-Solana:2013aba}
Jorge Casalderrey-Solana, Michal~P. Heller, David Mateos, and Wilke van~der
  Schee.
\newblock {From full stopping to transparency in a holographic model of heavy
  ion collisions}.
\newblock {\em Phys. Rev. Lett.}, 111:181601, 2013.

\bibitem{Garfinkle:2011tc}
David Garfinkle, Leopoldo~A. Pando~Zayas, and Dori Reichmann.
\newblock {On Field Theory Thermalization from Gravitational Collapse}.
\newblock {\em JHEP}, 02:119, 2012.

\bibitem{Garfinkle:2011hm}
David Garfinkle and Leopoldo~A. Pando~Zayas.
\newblock {Rapid Thermalization in Field Theory from Gravitational Collapse}.
\newblock {\em Phys. Rev.}, D84:066006, 2011.

\bibitem{Wu:2011yd}
Bin Wu and Paul Romatschke.
\newblock {Shock wave collisions in AdS5: approximate numerical solutions}.
\newblock {\em Int. J. Mod. Phys.}, C22:1317--1342, 2011.

\bibitem{Romatschke:2013re}
Paul Romatschke and J.~Drew Hogg.
\newblock {Pre-Equilibrium Radial Flow from Central Shock-Wave Collisions in
  AdS5}.
\newblock {\em JHEP}, 04:048, 2013.

\bibitem{Caceres:2014pda}
Elena Caceres, Arnab Kundu, Juan~F. Pedraza, and Di-Lun Yang.
\newblock {Weak Field Collapse in AdS: Introducing a Charge Density}.
\newblock {\em JHEP}, 06:111, 2015.

\bibitem{CaronHuot:2011dr}
Simon Caron-Huot, Paul~M. Chesler, and Derek Teaney.
\newblock {Fluctuation, dissipation, and thermalization in non-equilibrium
  AdS$_5$ black hole geometries}.
\newblock {\em Phys. Rev.}, D84:026012, 2011.

\bibitem{Danielsson:1999fa}
Ulf~H. Danielsson, Esko Keski-Vakkuri, and Martin Kruczenski.
\newblock {Black hole formation in AdS and thermalization on the boundary}.
\newblock {\em JHEP}, 02:039, 2000.

\bibitem{Danielsson:1999zt}
Ulf~H. Danielsson, Esko Keski-Vakkuri, and Martin Kruczenski.
\newblock {Spherically collapsing matter in AdS, holography, and shellons}.
\newblock {\em Nucl. Phys.}, B563:279--292, 1999.

\bibitem{Giddings:2001ii}
Steven~B. Giddings and Aleksey Nudelman.
\newblock {Gravitational collapse and its boundary description in AdS}.
\newblock {\em JHEP}, 02:003, 2002.

\bibitem{Lin:2008rw}
Shu Lin and Edward Shuryak.
\newblock {Toward the AdS/CFT Gravity Dual for High Energy Collisions. 3.
  Gravitationally Collapsing Shell and Quasiequilibrium}.
\newblock {\em Phys. Rev.}, D78:125018, 2008.

\bibitem{Baron:2012fv}
Walter Baron, Damian Galante, and Martin Schvellinger.
\newblock {Dynamics of holographic thermalization}.
\newblock {\em JHEP}, 03:070, 2013.

\bibitem{Steineder:2012si}
Dominik Steineder, Stefan~A. Stricker, and Aleksi Vuorinen.
\newblock {Holographic Thermalization at Intermediate Coupling}.
\newblock {\em Phys. Rev. Lett.}, 110(10):101601, 2013.

\bibitem{Steineder:2013ana}
Dominik Steineder, Stefan~A. Stricker, and Aleksi Vuorinen.
\newblock {Probing the pattern of holographic thermalization with photons}.
\newblock {\em JHEP}, 07:014, 2013.

\bibitem{Baier:2012ax}
Rudolf Baier, Stefan~A. Stricker, Olli Taanila, and Aleksi Vuorinen.
\newblock {Production of Prompt Photons: Holographic Duality and
  Thermalization}.
\newblock {\em Phys. Rev.}, D86:081901, 2012.

\bibitem{Baier:2012tc}
Rudolf Baier, Stefan~A. Stricker, Olli Taanila, and Aleksi Vuorinen.
\newblock {Holographic Dilepton Production in a Thermalizing Plasma}.
\newblock {\em JHEP}, 07:094, 2012.

\bibitem{Aparicio:2011zy}
Joao Aparicio and Esperanza Lopez.
\newblock {Evolution of Two-Point Functions from Holography}.
\newblock {\em JHEP}, 12:082, 2011.

\bibitem{Balasubramanian:2011ur}
V.~Balasubramanian, A.~Bernamonti, J.~de~Boer, N.~Copland, B.~Craps,
  E.~Keski-Vakkuri, B.~Muller, A.~Schafer, M.~Shigemori, and W.~Staessens.
\newblock {Holographic Thermalization}.
\newblock {\em Phys. Rev.}, D84:026010, 2011.

\bibitem{Balasubramanian:2010ce}
V.~Balasubramanian, A.~Bernamonti, J.~de~Boer, N.~Copland, B.~Craps,
  E.~Keski-Vakkuri, B.~Muller, A.~Schafer, M.~Shigemori, and W.~Staessens.
\newblock {Thermalization of Strongly Coupled Field Theories}.
\newblock {\em Phys. Rev. Lett.}, 106:191601, 2011.

\bibitem{Balasubramanian:2013oga}
V.~Balasubramanian, A.~Bernamonti, J.~de~Boer, B.~Craps, L.~Franti, F.~Galli,
  E.~Keski-Vakkuri, B.~Müller, and A.~Schäfer.
\newblock {Inhomogeneous holographic thermalization}.
\newblock {\em JHEP}, 10:082, 2013.

\bibitem{Balasubramanian:2013rva}
V.~Balasubramanian, A.~Bernamonti, J.~de~Boer, B.~Craps, L.~Franti, F.~Galli,
  E.~Keski-Vakkuri, B.~Müller, and A.~Schäfer.
\newblock {Inhomogeneous Thermalization in Strongly Coupled Field Theories}.
\newblock {\em Phys. Rev. Lett.}, 111:231602, 2013.

\bibitem{Balasubramanian:2012tu}
V.~Balasubramanian, A.~Bernamonti, B.~Craps, V.~Keränen, E.~Keski-Vakkuri,
  B.~Müller, L.~Thorlacius, and J.~Vanhoof.
\newblock {Thermalization of the spectral function in strongly coupled two
  dimensional conformal field theories}.
\newblock {\em JHEP}, 04:069, 2013.

\bibitem{Galante:2012pv}
Damian Galante and Martin Schvellinger.
\newblock {Thermalization with a chemical potential from AdS spaces}.
\newblock {\em JHEP}, 07:096, 2012.

\bibitem{Caceres:2012em}
Elena Caceres and Arnab Kundu.
\newblock {Holographic Thermalization with Chemical Potential}.
\newblock {\em JHEP}, 09:055, 2012.

\bibitem{Hubeny:2006yu}
Veronika~E Hubeny, Hong Liu, and Mukund Rangamani.
\newblock {Bulk-cone singularities \& signatures of horizon formation in
  AdS/CFT}.
\newblock {\em JHEP}, 01:009, 2007.

\bibitem{Erdmenger:2011jb}
Johanna Erdmenger, Shu Lin, and Thanh~Hai Ngo.
\newblock {A Moving mirror in AdS space as a toy model for holographic
  thermalization}.
\newblock {\em JHEP}, 04:035, 2011.

\bibitem{Erdmenger:2012xu}
Johanna Erdmenger and Shu Lin.
\newblock {Thermalization from gauge/gravity duality: Evolution of
  singularities in unequal time correlators}.
\newblock {\em JHEP}, 10:028, 2012.

\bibitem{Erdmenger:2011aa}
Johanna Erdmenger, Carlos Hoyos, and Shu Lin.
\newblock {Time Singularities of Correlators from Dirichlet Conditions in
  AdS/CFT}.
\newblock {\em JHEP}, 03:085, 2012.

\bibitem{Chesler:2012zk}
Paul~M. Chesler and Derek Teaney.
\newblock {Dilaton emission and absorption from far-from-equilibrium
  non-abelian plasma}.
\newblock 2012.

\bibitem{Chesler:2011ds}
Paul~M. Chesler and Derek Teaney.
\newblock {Dynamical Hawking Radiation and Holographic Thermalization}.
\newblock 2011.

\bibitem{Keranen:2014lna}
Ville Keranen and Philipp Kleinert.
\newblock {Non-equilibrium scalar two point functions in AdS/CFT}.
\newblock {\em JHEP}, 04:119, 2015.

\bibitem{Jarvela:2015zra}
Jarkko Järvelä, Ville Keränen, and Esko Keski-Vakkuri.
\newblock {Conformal quantum mechanics and holographic quench}.
\newblock 2015.

\bibitem{Ebrahim:2010ra}
Hajar Ebrahim and Matthew Headrick.
\newblock {Instantaneous Thermalization in Holographic Plasmas}.
\newblock 2010.

\bibitem{David:2015xqa}
Justin~R. David and Surbhi Khetrapal.
\newblock {Thermalization of Green functions and quasinormal modes}.
\newblock {\em JHEP}, 07:041, 2015.

\bibitem{Calabrese:2006rx}
Pasquale Calabrese and John~L. Cardy.
\newblock {Time-dependence of correlation functions following a quantum
  quench}.
\newblock {\em Phys. Rev. Lett.}, 96:136801, 2006.

\bibitem{Skenderis:2008dh}
Kostas Skenderis and Balt~C. van Rees.
\newblock {Real-time gauge/gravity duality}.
\newblock {\em Phys. Rev. Lett.}, 101:081601, 2008.

\bibitem{Skenderis:2008dg}
Kostas Skenderis and Balt~C. van Rees.
\newblock {Real-time gauge/gravity duality: Prescription, Renormalization and
  Examples}.
\newblock {\em JHEP}, 05:085, 2009.

\bibitem{D'Hoker:2002aw}
Eric D'Hoker and Daniel~Z. Freedman.
\newblock {Supersymmetric gauge theories and the AdS / CFT correspondence}.
\newblock In {\em {Strings, Branes and Extra Dimensions: TASI 2001:
  Proceedings}}, pages 3--158, 2002.

\end{thebibliography}

\end{document}